\documentclass{article}
\usepackage{latexsym,e-journal}

\usepackage[utf8]{inputenc}
\usepackage[english]{babel}

\usepackage{amssymb,amsfonts}

\usepackage[perpage,symbol*]{footmisc}
\usepackage[final]{graphicx}
\usepackage{pstricks}
\usepackage{cite, braket}

\usepackage[varg]{txfonts}

\begin{document}

\newcommand{\rum}{\rule{0.5pt}{0pt}}
\newcommand{\rub}{\rule{1pt}{0pt}}
\newcommand{\numtimes}{\mbox{\raisebox{1.5pt}{${\scriptscriptstyle \rum\times}$}}}
\newcommand{\numtimess}{\mbox{\raisebox{1.0pt}{${\scriptscriptstyle \rum\times}$}}}

\renewcommand{\refname}{References}
\renewcommand{\tablename}{\small Table}
\renewcommand{\figurename}{\small Fig.}
\renewcommand{\contentsname}{Contents}

\twocolumn[%
\begin{center}
\renewcommand{\baselinestretch}{0.93}
{\Large\bfseries Gauge Freedom and Relativity: A Unified Treatment of Electromagnetism, Gravity and the Dirac Field

}\par
\renewcommand{\baselinestretch}{1.0}
\bigskip
Clifford Chafin\\ 
{\footnotesize  \rule{0pt}{12pt}Department of Physics, 
North Carolina State University, Raleigh, NC 27695.
 E-mail: cechafin@ncsu.edu

}\par
\medskip
{\small\parbox{11cm}{%
The geometric properties of General Relativity are reconsidered as a particular nonlinear interaction of fields on a flat background where the perceived geometry and coordinates are ``physical'' entities that are interpolated by a patchwork of observable bodies with a nonintuitive relationship to the underlying fields.  
This more general notion of gauge in physics opens an important door to put all fields on a similar standing but requires a careful reconsideration of tensors in physics and the conventional wisdom surrounding them.  The meaning of the flat background and the induced conserved quantities are discussed and contrasted with the ``observable'' positive definite energy and probability density in terms of the induced physical coordinates.  
In this context, the  Dirac matrices are promoted to dynamic proto-gravity fields and the keeper of ``physical metric'' information.  Independent sister fields to the wavefunctions are utilized in a bilinear rather than a quadratic lagrangian in these fields.  This construction greatly enlarges the gauge group so that now proving causal evolution, relative to the physical metric, for the gauge invariant functions of the fields requires both the stress-energy conservation and probability current conservation laws.  
Through a Higgs-like coupling term the proto-gravity fields generate a well defined physical metric structure and gives the usual distinguishing of gravity from electromagnetism at low energies relative to the Higgs-like coupling.  The flat background induces a full set of conservation laws but results in the need to distinguish these quantities from those observed by recording devices and observers constructed from the fields.}}\smallskip
\end{center}]{%

\setcounter{section}{0}
\setcounter{equation}{0}
\setcounter{figure}{0}
\setcounter{table}{0}
\setcounter{page}{25}

\markboth{Clifford Chafin. Gauge Freedom and Relativity}{\thepage}
\markright{Clifford Chafin. Gauge Freedom and Relativity}

\markright{Clifford Chafin. Gauge Freedom and Relativity}
\section{Introduction}
\markright{Clifford Chafin. Gauge Freedom and Relativity}

{\looseness=+1
The theories (special and general) of relativity arose out of an extension of notions of geometry and invariance from the 19th century.  Gauge freedom is an extension of such ideas to ``internal'' degrees of freedom.  The gauge concept follow from the condition that quantities that are physically real and observable are generally not the best set of variables to describe nature.  The observable reality is typically a function of the physical fields and coordinates in a fashion that makes the particular coordinates and some class of variations in the fields irrelevant.  It is usually favored that such invariance be ``manifest'' in that the form of the equations of motion are evidently independent of the gauge.  Implicit in this construction is the manifold-theory assumption that points have meaning and coordinate charts do not.  We are interested in the largest possible extension of these ideas so that points themselves have no meaning and gauge equivalence is defined by mappings of one solution to another where the observers built of the underlying fields cannot detect any difference between solutions.  This is the largest possible extension of the intuitive notion of relativity and gauge.  It will be essential to find a mathematical criterion that distinguishes this condition rather than simply asserting some gauge transformation exists on the lagrangian and seeking the ones that preserve this.  This leads us to consider a more general ``intrinsic'' reality than the one provided by manifold geometry but, to give a unified description of the gravitational fields and the fields that are seen to ``live on top of'' the manifold structure it induces requires we provide an underlying fixed coordinate structure.  The physical relevance, persistence and uniqueness of this will be discussed, but the necessity of it seems unavoidable.  

}

Initially we need to reconsider some aspects of the particular fields in our study: the metric, electromagnetic and Dirac fields.  The Dirac equation is interesting as a spinor construction with no explicit metric but an algebra of gamma-matrices that induce the Minkowskii geometry and causal structure.  There are many representations of this but the algebra is rigid.  The general way to include spinors in spacetime is to use a nonholonomic tetrad structure and keep the algebra the same in each such defined space.  We are going to suggest an initially radical alteration of this and abandon the spinor and group notions in these equations and derive something isomorphic but more flexible that does not require the vierbein construction.  It is not obvious that this is possible.  There are rigid results that would seem to indicate that curvature necessitates the use of vierbeins \cite{Ma}.  These are implicitly built on the need for $\psi$ itself to evolve causally with respect to the physical metric (in distinction with the background metric).  We will extend the lagrangian with auxiliary fields so that this is not necessary but only that the gauge invariant functions of the \textit{collective} reality of these fields evolve causally.  This is a subtle point and brings up questions on the necessity of the positive definiteness of energy, probability, etc.\ as defined by the underlying (but not directly observable) flat space.    

Let us begin with a brief discussion of the Dirac equation and this modification.  The Dirac equation is the fundamental description for electrons in quantum theory.  It is typically derived in terms of causality arguments and the need for an equation of motion that is first order in time, as was Dirac's approach, or, more formally, in terms of representation theory of the Lorentz group.  These arguments are discussed many places \cite{Bjorken,Peskin, Weinberg}.  While this is a powerful description and has led to the first inclination of the existence of antiparticles, it has its own problems.  Negative energy solutions have had to be reconciled by Dirac's original hole theory or through the second quantization operator formalism.  Most are so steeped in this long established perspective and impressed by its successes that it gets little discussion.  

A monumental problem today is that of ``unification'' of quantum theory and gravity.  There are formal perturbative approaches to this and some string theory approaches as well.  In quantum field theory we often start with a single particle picture as a ``classical field theory'' and then use canonical quantization or path integral methods.  For this reason, it is good to have a thorough understanding of the classical theory to be built upon.  We will show that, by making some rather formal changes in traditional lagrangians, some great simplifications can result.  The cost is in abandoning the notions that the fields corresponding to nature are best thought of as evolving on the ``intrinsic'' geometry induced by a metric and that spacetime is a locally Lorentzian manifold.  In place of this is a trivial topological background and a reality induced by fields which encodes the observable reality and apparent coordinates (induced by collections of objects) and metrical relationships in a non-obvious fashion.  Usual objections to such a formalism in the case of a gravitational collapse are addressed by adherence to the time-frozen or continued collapse perspective.

A main purpose of this article is to illustrate an alternate interpretation of the Dirac equation.  In the course of it, we will make gravity look much more like the other bosonic fields of nature and give a true global conservation law (that is generally elusive in GR).  Our motivation begins with a reconsideration of the spinor transformation laws and the role of representation theory.  This approach will greatly expand the gauge invariance of the system.  In place of the metric $g_{\mu\nu}$ as the keeper of gravitational information, we will let the $\gamma$ matrices become dynamic fields and evolve.  Our motivation for this is that, for vector fields, the metric explicitly appears in each term and variation of it, gives the stress-energy tensor.  The only object directly coupling to the free Dirac fields is $\gamma$.  Additionally, $\gamma^{\mu}$ bears a superficial resemblance to $A^{\mu}$ and the other vector bosons.  Since $g\sim \gamma\gamma$ we might anticipate that the spin of this particle is one rather than two as is for the graviton theories which are based explicitly on $g^{\mu\nu}$.  It is because we only require our generalized gauge invariant functions to obey causality and that these conserved quantities, while exact, are not directly observable so do not have to obey positive definiteness constraints that this approach can be consistent.  

We will be able to show that this construction can give GR evolution of packets in a suitable limit and obeys causal constraints of the physical metric.  It is not claimed that the evolution of a delocalized packet in a gravitational field agrees with the spinor results in a curved spacetime.  This will undoubtably be unsatisfactory to those who believe that such a theory is the correct one.  In defense, I assert that we do not have any data for such a highly delocalized electron in a large nonuniform gravitational field and that the very concept of spinor may fail in this limit.  As long as causality holds, this should be considered an alternate an viable alternative theory of the electron in gravity.  The purely holonomic nature of the construction is pleasing and necessary for a theory built on a flat background.  A unification of gravity in some analogous fashion to electroweak theory would benefit from having a its field be of the same type.  One might naturally worry about the transformation properties of $\psi_{a}$ and $\gamma^{\mu}_{ab}$ in this construction.  Under coordinate transformations of the background, $\psi_{a}$ behaves as a scalar not a spinor and $\gamma^{\mu}_{ab}$ is a vector.  One should not try to assign to much physical meaning to this since these transformations of the structure are passive.  Active transformations where we leave the reality of all the surrounding and weakly coupled fields the same but alter the electron of interest can be manifested by changes in both $\psi$ and $\gamma$ (and $A$) so that the local densities and currents describing it are boosted and those of the other fields are not.  The usual active boost $\psi'_{b}=S(\Lambda)_{ba}\psi_{a}$ is included as a subset of this more general gauge change.  

There has been work from the geometric algebra perspective before \cite{Hestenes} in trying to reinterpret the Dirac and Pauli matrices as physically meaningful objects.  Since the author has labored in isolation for many years searching for a physical meaning for the apparent geometric nature of physical quantities this did not come to his attention until recently.  However, there are significant differences in the approach presented here and the easy unification with gravity that follows seems to depend on abandoning group representation theory in the formulation.  Most importantly, one has a new notion of gauge freedom as it relates to the reality expressed by particle fields (i.e.\ the full gauge independent information associated with it).  Coupling destroys the ability to associate the full ``reality'' of the electron with the wavefunction.  We will see that this can get much more entangled when one includes gravity and, with the exception of phase information, the only consistent notion of a particle's reality comes from the locally conserved currents that can be associated with it.  Here will involve multiple field functions not just $\psi_{a}$ as in the free particle case.  

The dominant approaches to fundamental physics has\linebreak
 been strongly inspired by the mathematical theory of manifolds where a set of points is given a topology and local coordinate chart and metric structure.  The points have a reality in this construction and the charts are grouped into atlases so that coordinates are ``pure guage'' and no physical reality is associated with them.  We frequently say that the invariance of the field's equations requires that we have a metric invariant action be a scalar.  It can be shown somewhat easily \cite{Chafin-nonloc} that this is not true and that most lagrangians that give many common (local) field equations are neither invariant nor local.  In the following we enlarge the class of physically equivalent fields to the set of fields that evolve in such a fashion where the ``observers'' built from the fields cannot distinguish one description from another.  This includes simple spacetime translations of a flat space of the entirety of fields and far more general deformations of the fields which do not preserve the underlying set of points.  

The underlying space is chosen trivially flat with the \hyphenation{Lor-ent-zian} $\eta^{\mu\nu}$ metric.  This begs the question of how general curved coordinates resulting from the effective curvature induced by the field $g^{\mu\nu}(\gamma)$ relate to it and how the causally connected structure induced by the fields evolves through this flat background.  In this picture the ``physical coordinates'' seen by\linebreak
 observers are measures induced by ``candles,'' specifically\linebreak
 highly independent localized objects and radiators, that induce his perception of his surroundings.  Clocks are induced by atomic oscillations and other local physical processes.\linebreak
 Collective displacements and alterations of the fields on the underlying flat space that preserve the preserved reality are considered alternate representations of the same physical reality rather than an active transformation of it to a new and distinct one, as one would expect from the usual manifold founded perspective.  

At the foundations of manifold inspired physics are tensors and their transformation rules under coordinate changes.  In this case we have little interest in the transformations with respect to the underlying flat space and all fields are treated as trivial tensors with respect to it.  The interesting case of apparent curvature must then be measured with respect to these local candles.  The vector properties of functions of a field, like the current $j_{(0)}^{\mu}=\bar\psi_{(0)}\gamma^{\mu}\psi_{(0)}$, are then the collective result of active transformations of the $\psi_{(i)}$, $\gamma$ and underlying coordinates that leave the nearby candles' (labelled by $i$) gauge invariant features unchanged and a transformation of the field $\psi_{(0)}$ so that the resulting current $j_{(0)}$ appears to move through a full set of Lorentz boosts and rotations relative to measurements using these candles.  

This is a significant departure from the usual geometry inspired approach.  Not surprisingly many formulas will appear (deceptively) similar to usual results despite having very different meaning since they will all be written with respect to the underlying flat structure not some ``physical coordinates'' with respect to some fixed point set induced by the candles.  The mystery of how we arrive at a geometric seeming reality and at what energy scale we can expect this to fail is a main motivation for this article.  Conservation laws follow from the usual ten Killing vectors of flat space but the meaning of these conservation laws (and their form in terms of observable quantities) is unclear.  Even the positive definiteness of quantities like energy and mass density are not assured and failure of them do not carry the same consequences as in usual metric theories.  The symmetry responsible for mass conservation is the same one as for probability so such a situation raises more questions that must be addressed along the way.  We have been nonspecific about the details of what determines equivalent physical configurations.  Aside from the geometry induced by candles the gauge invariant quantities that we presume are distinguishable by observers are those induced by conserved currents such as mass and stress-energy.  It is not obvious why such should be the case.  A working hypothesis is that all observers are made up of long lasting quasilocalized packets of fields that determine discrete state machines and these are distinguished by localized collections of mass, charge and other conserved quantities.  

In this article we only discuss these as classical theories in a 4D spacetime.  Of course, the motivation is for this to lead to a general quantum theory.  There is a lot of work on reinterpretation of quantum theory as a deterministic one.  Everyone who works on this has his favorite approach.  The author here is no exception and has in mind a resolution that is consistent with the theory in \cite{Chafin} that gives QM statistics assuming that particular far-from-eigenstate wavefunctions describe classical matter that arise in an expanding universe with condensing solids.  The motivations behind the following constructions is not just to get some insight on unification but to take steps to resolve some of the fundamental contradictions of quantum field theory, such as Haag's theorem, and to give a solid justification for the calculations of field theory that have been successful.  

The structure of the article will be as follows.  Invariance and the nature of causality are discussed and contrasted with the usual flat background approach in \S2.  This is especially subtle since the ``physical'' metric, reality and coordinate features are encoded in this construction in nonobvious ways, the gauge group is large and some conserved quantities and expected positive definiteness of quantities can change without altering the physically observable results.  Next we will elaborate in \S3 on the transformation properties of the fields and promotion of the gamma matrices to holonomically described proto-gravity fields in causally consistent manner and in \S4 give a discussion on the ``reality'' induced by fields.  In \S5 we modify the Dirac lagrangian with an auxiliary field $\phi$ to replace the awkward $\bar{\psi}={\psi}^{*}\gamma^{0}$ with its extra $\gamma^{0}$ factor uncontracted in any tensorial fashion, and demonstrate causality of the gauge invariant functions of the field.\footnote{We typically vary $\psi$ and $\psi^{*}$ independently in the lagrangian to get equations of motion but then constrain them to be so related (though we should show this constraint is propagated as well). Here we make no such restriction and allow $\psi$ and $\phi$ to be independent fields with no constraints on the initial data.  In the flat space case, the case of $\phi=\gamma^{0}\psi^{*}$ gives the usual results and shows many other cases (i.e.\ $\psi,\phi$ initial data pairs) are gauge related to this.  In the case of a nontrivial gravity field, we allow the possibility that no such mapping may exist.} 
In \S6, a sister field to $\gamma$ is introduced that allows a similar lagrangian for the proto-gravity fields (when a Higgs-like construction is used) as for the electromagnetic field and that gives General Relativity in a suitable limit.  This similarity suggests a pairing of the electromagnetic and proto-gravity fields in a manner reminiscent of the electroweak theory.  \S7 gives a discussion of the global conservation laws that arise due to symmetries of the flat background.




\markright{Clifford Chafin. Gauge Freedom and Relativity}
\section{Roles of invariants in physics}
\markright{Clifford Chafin. Gauge Freedom and Relativity}

The mathematical theory of invariants arose in the 19th century and the intuition derived from them made a physical appearance with the work of Mach \cite{Mach} and Einstein \cite{Einstein}.  Since then they have played a preeminent role both in formulating theory and solving particular problems.  The geometrodynamic approach to General Relativity is to assume some underlying geometry that is locally special relativity and posit that this geometric structure and its associated transformation laws are the natural way to look at the world.  ``Flat background'' approaches are generally to look at small post-Newtonian corrections to the universe for nearly flat spaces where gravity is playing a small role \cite{MTW}.  In more dramatic configurations this formalism seems hopelessly flawed.\linebreak
  Wormholes are topologically forbidden from such a description.  Black holes with their singularities have infinite metric curvature at the center and the interior of the event horizon causally decouples in one direction from the exterior.  

There is an old and out-of-favor view of black holes that goes back to Oppenheimer \cite{Oppenheimer} whereby the infalling matter gets redshifted to an effective asymptotic standstill so that no singularity or horizon ever forms.  This is often called the ``time-frozen'' picture.  For many this is considered equivalent to lagrangian evolution where the particles fall in finite proper time to the center.  It is usually neglected that this implies a transfinite amount of external observer time must elapse for this to occur.  This implies that we have assumed that in the entirety of external observer time, no collective action occurs to interfere with black hold formation before the event horizon forms.  Furthermore, an infalling pair of charges on opposite nodes will be seen as a dipole field for all future time in the time-frozen case.  The lagrangian approach would suggest that these fall to the center and form a spherically symmetric charge distribution as suggested by the ``no-hair'' conjecture.  This latter picture has no physical relevance for the external observers, so the author is firmly in the time-frozen camp.  

The importance of this point of view is that there are no exotic topologies to get in the way of assuming that one has a flat background.  The ``geometric'' aspects of gravity are some yet to be explained feature of a field that evolves in an equivalent fashion to all the other fields of nature.  Let us now take the point of view that there is a flat background and, rather that looking at perturbations of it as $g^{\mu\nu}=\eta^{\mu\nu}+\tilde h^{\mu\nu}$, the field $h^{ab}$ sits on top of it and is coupled to the other fields, including the kinetic terms, in the fashion of a metric.  Let this background have the flat space metric $\eta^{\mu\nu}$ so that coupling, for the electromagnetic case, is of the form
$$\mathcal{L}=\left(\partial_{\alpha}A_{\beta}-C_{\alpha\beta}^{\gamma}A_{\gamma}\right)\, h^{\alpha\alpha'}h^{\beta\beta'} \left(\partial_{\alpha'}A_{\beta'}-C_{\alpha'\beta'}^{\gamma'}A_{\gamma'}\right),$$
where the
connection-like $C$ tensor is yet to be defined.  
Importantly, these are \textit{not} considered to be indices that transform as co and contravariant tensors under the metric $h$.  All the objects here are flat space $\eta$-tensor objects.  This seemingly bazaar construction gives causal cones for the evolution that are not the flat space cones defined by $\eta^{\mu\nu}$.  The coordinate labels $\hat t, \hat x, \hat y, \hat z$ give coordinate directions.  We expect that the $(x,y,z)$ set are $h$-spacelike in the sense that $h^{ij}u_{i}u_{j}>1$ for all $u$ in the span of $\hat x, \hat y, \hat z$.  The forward timelike direction has a positive projection on $\hat t$ even if the cone is so tilted that $h^{tt}>0$.   Thus it gives a positive evolution direction for a future on the background.

In general, any reasonable equation of motion for $h$ should preserve this set of conditions and evolve in our coordinate time variable $t$ for all values.  In the case of black hole formation the metric tends to asymptotically converge on a degenerate state leading to a set of equations that are very ill-conditioned.  How to treat this situation numerically is still unclear but the presence of a flat $\eta$-background means that we have a full set of conservation laws so these may provide an avenue to evolve without such problems \cite{Chafin-BH}.  We will not be answering the question of general persistence of evolution of the equations as it seems to be a very hard problem (as most nonlinear PDE solution existence problems are) but it is very important.  Failure of this to hold would be destructive to such a theory.  It is taken as an article of faith that such a set of initial data can be evolved for all coordinate time with time steps taken uniformly at all locations.  In other words, cones may narrow and tilt but they will never intersect with our spatial coordinate slices.

The role of gauge invariance in physics is analogous to an equivalence class in mathematics.  In mathematics we have some set of structures we wish to preserve and there can be classes of elements that act the same under them.  In physics, we may have a set of fields that evolve under the equations of motion in such a way that there are classes that retain some set of properties under evolution.  We usually describe the set by a gauge transformation that joins each subclass.  It is not clear that nature is really blind to which element of the class we are choosing.  One could choose a representative element and claim that this is the ``correct'' one and be no worse for it.  In the case of the Dirac field $\psi$ and the electromagnetic field $A$ each has a set of gauge transformations as free fields.  The Dirac field has only a global phase transformation however, when coupled to the electromagnetic field, it acquires some local gauge freedom $A\rightarrow A+\nabla\chi$ in that the phase $\varphi\rightarrow \varphi-\chi$.  This is what we mean by ``promoting'' a global to a local symmetry.

In the following we will replace the quadratic lagrangian with a bilinear one by replacing $\bar{\psi}={\psi}^{*}\gamma^{0}$ with a new field $\phi$.\footnote{Such a construction also introduces a large set set of nonlocal conservations laws. \cite{Chafin-nonloc}}  This is the motivation for the title.  We are really only abandoning $\gamma^{0}$ in this sense as a factor in defining $\bar\psi$.  The fields $\gamma^{\mu}$ are all retained as what might be loosely called a ``spin 1'' encoding of the gravitational field.  We now need to ask what are the physically distinguishable states of the system.  It is natural to argue that the conserved quantities give the only unambiguous physical quantities that we can distinguish.  Phase is complicated in that it gives current and relative cancellation due to interference.  One can define a $\psi$ by the mass density $\rho$ and the current $j$.  When the density is over a compact set this is enough to fix the phase up to a constant.  For our new set we will have conservation laws that depend on $\psi$, $\phi$ and $\gamma$.  The $\gamma^{0}$ is still present but now a dynamical field.  This trio of fields now collectively determines the conserved currents.  Naturally this is a massive expansion of the gauge group.  In the ``flat space'' case we can choose $\gamma^{\mu}$ to be the Dirac matrices in some representation and $\phi=\gamma^{0}\psi^{*}$ and obtain the usual Dirac results.

The Noether charge symmetries here correspond to spacetime symmetries and phase transformations.  When we consider the quantum analogs of such fields the importance of positive definite norm is important.  This is because it is given the role of a probability for a measurement so must be positive definite and normalizable.  This fails in the classical theory of Dirac particles but is ``fixed up'' in the quantum field theory by choices for the commutation relations of the operators and their action on the vacuum ground state (as with the Gupta-Bluer formalism \cite{Schweber}).    In this classical theory we are not necessarily concerned with this for this reason but the same symmetry generates mass and charge conservation so it still is important.  Interestingly, this symmetry holds in curved space as we propagate hyperbolic spacelike slices even when there is no spacetime symmetry.  

One way the Dirac field is incorporated into curved spacetime is to fix $\gamma^{\mu}$ set to be a particular representation and use vierbein fields (tetrad formalism).  This preserves the desired norm properties above and ensures local packets move correctly.  There is little choice in this approach if one is to use wavefunction evolution from a quadratic lagrangian \cite{Ma}.  To be fair, no one knows what the evolution of an electron is on such scales.  We expect packets to move along geodesics but if some negative norm or mass density entered we then must defer to experiment to validate or reject this.  The probabilistic interpretation seems hopeless but consider that true ``observers'' as machines that measure the results are themselves built from such fields.  If quantum evolution is a deterministic feature as decoherence advocates suggest, then the probability is unity by the evolution and a change in positive definite norm means that the action of our measurement devices must obey a modified rule that preserves this.  This should be kept in mind when we consider questions about the conserved quantities.  Negative energy and mass regions of quantum bodies in highly curved regions my not be forbidden by nature as much as we forbid it by our assumptions about the essential meaning of such quantities.  

For evolution on such a flat $\eta$-background that mimics gravity, we must then ask what kinds of transformations correspond to the general coordinate transformations we are used to in GR.  Firstly, just as information has come to be considered a physical state in quantum information theory, coordinates and time should be thought of as physical conditions given by the kinds of candles afforded by local atoms and clusters that triangulate our spacetime.  We may as well think of ``physical coordinates'' (i.e.\ non $\eta$-background coordinate changes) as made of material bodies that are small enough to give insignificant perturbations to the general dynamics.  To actively boost to another RF (reference frame) we consider a local current relative to some other standard currents that define the frame and choose the new current so the relative local motion matches.  To passively boost to another RF we consider a transformation of the underlying $\eta$-background coordinates.  Since the physically causal light cones induced by $h^{\mu\nu}$ in its coupling to the other fields $A$, $\psi$, etc.\ are not the cones induced by $\eta$ we must take care to maintain the $\hat t$-forward direction of the cones under such changes.  The tensor field constructions made with the usual forms $\bar \psi \gamma_{D}^{\mu}\psi$, etc.\ will now be of the form $j^{\mu}=\phi \gamma^{\mu}\psi$ so that their transformation properties under $\eta$-background coordinate changes are tensorial.  This is, however, not very interesting because it does not relate to our physical observers and their physical coordinates that relate to the function $h^{\mu\nu}$.  Many active transformation of the field trio $\phi,\gamma,\psi$ give the same boosted current.  If we make the change purely with $\gamma$ and assume our metric function $h^{\mu\nu}$ is built from them, this will change other terms in the equations of motion.  

There remains the many possibilities of transforming the pair $\psi,\phi$ to give a new current function without altering the local observed geometry.  Passive transformations based on\linebreak
 allowable background coordinate changes can be done by\linebreak
 changing the $\eta$-background coordinates or altering the fields $\psi,\phi$ in a manner that gives a shifted (on the background coordinates) set of currents and conserved densities that evolve in an isomorphic fashion to the original fields.  The possibility of having shifted and deformed sets of fields on the background space with the same observable reality is a novel extension over the manifold approach where the points have reality and we assign and transform fields there based on coordinate changes and other gauges.  It is analogous to having a set of fields on $\mathbb{R}^{4}$ and shifting the set by a 4-vector $v^{\mu}$ to give a new equivalent universe of solutions in the equivalence class; an obviously true equivalence that is not present by positing a manifold with fields.  We now allow this full set of equivalent representations of such a universe.

\markright{Clifford Chafin. Gauge Freedom and Relativity}
\section{Transformation rules}
\markright{Clifford Chafin. Gauge Freedom and Relativity}

The theory of spinors arose naturally out of Dirac's algebraic attempts to reconcile causality with the first order equations that seem to describe nonrelativistic electrons.  Interestingly, Schr\"{o}dinger originally attempted the, later named, Klein-Gordon equation to describe electrons but could not get the fine structure right \cite{Weinberg}.  He settled on a diffusion-like equation that was first order in time and second order in spatial derivatives.  Pauli adapted it to include spin but, as for most such equations, signal propagation speeds diverge.  Dirac introduced a pair of spinors and a linear first order operator that when ``squared'' gave the Klein-Gordon equation for each component, thus ensuring causality.  

His treatment introduces a set of $\gamma^{\mu}_{ab}$ matrices that are considered fixed and constitute representations of the SL$(2,\mathbb{C})$ group which is a two-fold covering group of the SO$^{+}(3,1)$ group.  More explicity, this gives a map of complex valued bi-spinors ${a\choose b}{c\choose d}$ to real 4-vectors so that each 4${\times}$4 complex matrix action corresponds to a Lorentz transformation and compositions among these is preserved by this mapping.  In the humblest of terms, we can decompose a general free state $\psi_{a}$ into a basis of free progressive wave solutions $e^{ik_{\mu}x^{\mu}}{u}_{a}(k)$ where we can define a general Lorentz transformation $\Lambda^{\mu'}_{\nu}$ through the coordinate \textit{and} algebraic action $S(\Lambda)_{ab}\psi_{b}(\Lambda x)$.  We define this action so that the current $j^{\mu}$ is transformed by a boost and interpret it as the actively boosted free plane wave of positive energy.  Note that $S(\Lambda)_{ab}\psi_{b}(x)
\ne\psi_{a}(\Lambda x)$.

The Dirac lagrangian has a (seemingly) symmetric form
\begin{equation}
\mathcal{L}_{D}=i\bar{\psi}\gamma^{\mu}\partial_{\mu}\psi-m\bar{\psi}\psi,
\end{equation}
where $\bar{\psi}={\psi}^{*}\gamma^{0}$.  This inconvenient $\gamma^{0}$ is generally considered necessary to give Lorentz invariance.  We can see that without it we would get inconsistent equations of motion for $\psi$ and $\psi^{*}$ if we vary them independently.

The operator $S(\Lambda)_{ab}$ performs a transformation of $\psi_{a}$ so that the lagrangian is invariant and the resulting current is boosted as 
\begin{equation}
\begin{array}{l}
\displaystyle
j'^{\alpha}(x')=\bigl({\psi}'(x')^{*}\gamma^{\alpha}{\psi}'(x')\bigr)\\[6pt]
\displaystyle
\phantom{j'^{\alpha}(x')}
=\bigl({(S\psi(x))}^{*}\gamma^{\alpha}{S\psi}(x)\bigr) =\bigl(\psi(x)^{*}S^{*}\gamma^{\alpha}{S\psi}(x)\bigr)\\[6pt]
\displaystyle
\phantom{j'^{\alpha}(x')}=\left(\psi(x)^{*}\gamma^{'\alpha}{\psi}(x)\right)= \Lambda^{\alpha}_{\beta}\left({\psi}^{*}(x)\gamma^{\beta}{\psi}(x)\right)\\[6pt]
\phantom{j'^{\alpha}(x')}=\Lambda^{\alpha}_{\beta}j^{\beta}(x).
\end{array}  
\end{equation}
The Dirac theory allows us to think of the complex 4-spinors $\psi_{a}$ at each point as indicating the local direction of the local current of the particle corresponding to it.  To achieve this it has been necessary to introduce negative energy solutions.  
The negative energy solutions are reinterpreted as positrons and given a positive mass through the details of canonical quantization since they are generally deemed undesirable. 
One reason to reconsider this point is that net positive energy initial data may maintain this property and negative energy states do not necessarily provide an avenue for some subset of the space to fall to negative infinite energy at the expense of heating the rest of the system.  Such a result would depend on the details of the coupling and dynamics.  Local net negative energy density in solutions arising from positive local energy physically arising states would produce problems but it is not clear that this ever arises except in extreme cases where pair production becomes available.  

Other conservation laws such as the conservation of probability (which arise from the same global phase symmetry that give mass and charge conservation) have similar problems.  In an ``emergent'' theory of quantum measurement we do not need a probability operator (or any operators at all).  The probabilities arise from measurements with the kinds of macroscopic yet still quantum mechanical matter that constitutes the classical world \cite{Chafin}.  In this approach, the initial data and evolution equations generate their dynamics in a deterministic fashion and the probabilistic features arise from the long lived partitioning of the classical world into subsets indexed by the delocalized objects that interact with it.  Details of when this is a consistent procedure are discussed in ref.\ \cite{Chafin}.   For this reason, we do not seek to validate or build upon arguments that start with an ``interpretation'' of particular expressions since we ultimately expect the evolution and interactions to independently determine the expressions that give all observable results.  

One of the frustrating aspects of the Dirac equation as it stands is that it is not clear how we should alter its form in general coordinates.  One can use the local frame approach and assume the Dirac matrices are members of the same representation in each one.  A spinorial connection then indicates how nearby spinors are related as a consequence of geometry.  If we allow the matrices to become functions of space and time with only the spacetime indices changing this gives a simple approach but then it is not clear how we recover local Klein-Gordon (KG) evolution of each component and what the locally boosted fields should be.  If we continue with the spinor approach and let the $\gamma^{\mu}(x)$ matrices be fixed and alter the spinor fields instead then we need a transformation that is a kind of ``square root'' of the Lorentz vector transformation.  This is how we get the actively boosted solutions  in flat space.  In curved spacetime, there is no global notion of a boost so the former perspective seems more valuable.  Ultimately, we specify a configuration by the spacetime metric and the fields on it but the metric will be a function of the $\gamma^{\mu}$ matrix fields (and some associated dual fields) that only give geodesic motion below some energy bound.

In the early days of the Dirac equation, interpretations have evolved from a proposed theory of electrons and protons to that of electrons and positrons with positrons as ``holes'' in an infinitely full electron ``sea'' to that of electrons with positrons as electrons moving ``backwards in time.''  The first interpretation failed because the masses of the positive and negative energy parts are forced to be equal.  The second was introduced out of fear that the negative energy solutions of the Dirac equations would allow a particle to fall to endlessly lower energies.  The last was introduced as a computational tool.  The negative mass solutions were to be reinterpreted as positive mass with negative charge.  Necessary computational fixes associated with this idea are subtly introduced through the anticommutation relations used in the field theory approach to fermions and the properties of the supposed ground state \cite{Schweber}.  If we are going to seek a classical field theory approach to this problem we need another mechanism.   

For the moment, we assume the $\gamma$ matrices are those of the Dirac representation.  Standard treatments allow any selection of 4${\times}$4 matrices that represent the  SO$^{+}$(3,1) group.  Here we choose a specific representation because we are going to let the $\gamma$'s be fields and let these other choices be a kind of gauge freedom until some interaction restricts us to a specific subset. 
The Dirac lagrangian has a (seemingly) symmetric form
\begin{equation}
\mathcal{L}_{D}=i\bar{\psi}\gamma^{\mu}\partial_{\mu}\psi-m\bar{\psi}\psi
\end{equation}
where $\bar{\psi}={\psi}^{*}\gamma^{0}$.  This is generally considered necessary to give Lorentz invariance.
The Dirac matrices satisfy the condition 
\begin{equation}\label{gamma}
\{\gamma^{\mu},\gamma^{\nu}\}=-2\eta^{\mu\nu},
\end{equation}
where $\eta=${Diag}$(-,+,+,+)$.  This suggests that we could view the metrical properties of the space as encoded in $\gamma$ rather than invoking a metric $\eta$.  The metric has ten independent parameters at each point and $\gamma$ has $4\times10$ or $4^{3}$ parameters, depending on chosen symmetry constraints but we need to satisfy $4^{4}$ equations.  If we trace the suppressed spin indices then there are only 10 equations and a general metric can be encoded in the $\gamma^{\mu}$ set.  However, eqn.\ \ref{gamma} is the identity we require to convert the Dirac equation into a KG one that demonstrates causality in each component.  This is a loose end in deriving geodesic motion for a packet to show that we get observed motion in the classical GR limit and an important consideration in what follows.  

In anticipation of a future unification theory one cannot help but notice the greater similarity of $\gamma^{\mu}_{ab}(x)$ to $A^{\mu}(x)$ and the other vector boson fields than any of these to the metric $g_{\mu\nu}$.  For now we simply leave this as constant but accept that it can have its own transformation properties as a one-vector.  In contrast, all the ``spinor'' labels are considered as having only scalar transformation properties.  The bispinors $\psi_{a}$ now transform as scalars.  To emphasize their new properties and that they still have a collective reality as a four-tuple of functions we term it a ``spinplet.''  The mixed objects $\gamma^{\mu}_{ab}$ we consider a vector object with extra labels and, by analogy, label it a ``vectorplet.''  

There are some surprising implications of this.  The equations are unchanged but the transformation properties are now different.  Since the $\gamma^{\mu}_{ab}$'s can vary with position, we expect a much larger equivalence class of electron-gravity field pairs, $\{\psi,\gamma\}$, that correspond to the same underlying reality.  We can boost the system by $\Lambda_{\alpha}^{\mu}\gamma^{\alpha}$.  This gives the same $\psi_{a}$ fields at every point but the physically measurable $j^{\nu}$ currents are altered.  Of course we still have the traditionally boosted solutions $S(\Lambda)\psi^{(0)}(\Lambda x)$ that have this same current so we have a degeneracy in the pairs $(\Lambda\gamma,\psi^{(0)})$ and $e^{i\phi}(\gamma, S(\Lambda)\psi^{(0)})$ and all other states with the same current and net phase.  This is not the result of a discrepancy in the active vs.\ passive coordinate transformations we observe in a fixed representation but an additional degeneracy in the equivalent physical descriptions.  We have only used the current $j^{\mu}$ to distinguish states and we expect that there will be some other conserved quantities, like stress-energy, that will physically subdivide this set into distinct equivalence classes.  Since there are so many degrees of freedom in the set of $\gamma^{\mu}_{ab}(x)$'s we anticipate that the set is still significantly enlarged.

\markright{Clifford Chafin. Gauge Freedom and Relativity}
\section{Reality and gauge}
\markright{Clifford Chafin. Gauge Freedom and Relativity}
\label{reality}
 The AB effect gives a simple example of how the ``reality'' of an electron is not sufficiently described by the wavefunction of the electron itself.  In this case, the current is a function of both $\psi$ and $A$ as $J=i\hbar\nabla \psi+eA$.  This construction is useful in sorting out various apparent contradictions in electromagnetism.  If we want to investigate the radiation reaction or questions of ``hidden momentum'' \cite{Jackson, Rohrlich} one can build a packet that spreads slowly compared to the effects of external fields and see how the self field and lags contribute to the actual motion.  The power of it is that there is no ambiguity in the gauge as for a hodge-podge lagrangian like $\frac{1}{2}mv^{2}+jA-\frac{1}{4}FF$ \cite{LL} because the physical current of a packet is the gauge invariant $J$ not the naive $j=mv$.  The AB effect seems like a topological effect because it is viewed through the lens of $\psi$ being the pure descriptor of the reality of the electron and as a stationary effect.  In driving a solenoidal current to create a circulating $A$ field we accelerate $J$ with a transient circulating $E$ field.  Part of the current is made up of the phase gradient of $\psi$ and part from $A$ itself.  The field and the acceleration moves outwards from the current source at the speed of light and the resulting equilibrated current becomes a function of the final magnetic flux.  This circulating current must gain all of its curl from $A$.  The $\psi$ can only contribute to an irrotational flow so general charge packet motion requires a contribution from $A$.  
This suggests we might generally want a more nuanced distinction of particle reality than merely a function of each individual field in a lagrangian that has been nominally assigned to the particle type alone.  


In flat space without gravity or interactions, we can consider packets of fields that are widely separated based on type.  These can then evolve separately and the type of field and the reality implied by it are synonymous.  There can still be some gauge freedom but the packets and any interesting properties that one might observe are contained in the same support.  The observables are, at best, the gauge invariant properties such as stress-energy or current.  Allowing interactions, this reality gets complicated in two ways.  Firstly, the conserved currents may now involve aspects of more than one kind of field and second, there are now constraints that must be obeyed.  These are generally defined by elliptic PDEs such as $\nabla\cdot E=\rho$ that are propagated by the dynamic equations.\footnote{This is purely a classical theory of delocalized fields so we do not have the problem of ``self-energy'' or the ``particle not feeling its own fields.''   In the many body case, the fields presumably are made of many constituent ones with only the ``center of mass'' motion as visible to us.  This allows us to have a wavefunction of a charged particle that does not spread under the influence of the field generated by it, as in the classical particle case \cite{Rohrlich}.  However, the self force and momentum are subtle concepts in that such a composite charge must have both $m_{bare}$ and $m_{em}$ components.  Only $m_{bare}$ is localized and $m_{em}$ is spread over the range the static fields. The contribution to the electromagnetic momentum in $Ma=(m_{bare}+m_{em})a=F_{ext}$ in the force law is actually provided by a self field of the radiation field traversing the support of the charge.  }

If we now include gravity in the form of a $\gamma_{\mu}$ field that has some gauge freedom that mixes with the reality of the wavefuction $\psi$ then we cannot make the above separation.  The gravitational field is everywhere so no isolation of packets is possible.  The reality of the electron is now a function of $\psi$ and any $\gamma$-like fields that have global extent.  This is in contrast with the case where the gravitational information is completely specified in the $g^{\mu\nu}$ field.  Since this has no gauge freedom beyond that of coordinate changes, the packet motion of a wavefunction is affected by it yet the reality of the electron is still entirely determined by the values of $\psi$ in the packet itself.  

For the case where multiple fields determine a single reality, when is it really viable to call one set of quantities the ``electron current'' versus some combination of quantities that strictly depend on multiple types of fields?  In the case of the Dirac and electromagnetic field (in flat space with constant $\gamma$ matrices), the density of the field is only a function of $\psi$ so that we have at least one component of the 4-current that is entirely specified by the wavefunction.  This allows us a uniquely associate $j^{0}$ with the electron field $\psi$ and so call it the ``electron-density.''  The stress-energy terms similarly have $T^{00}$ as a simple function of $\psi$ alone.  If every conserved quantity can be associated this way, we have a well-defined mapping between the fields and conserved quantities.  If we are interested in more exotic lagrangians than can be formed by the ``minimal'' prescriptions from the free quadratic cases, we will need to be mindful of the possibility that the currents may not necessarily be so associated with one particular field.  

Although this discussion may feel somewhat pedantic, it is important to make this distinction  and not get trapped in the vague lore that sometimes accompanies discussions in physics.  For example, it is often said that we must have ``manifestly invariant'' lagrangians to get relativistically consistent results.  This is not true not only in the obvious sense that they can be rearranged in a nonobvious invariant form.  One can conceivably write down a set of fields that gives a class of solutions whereby the degrees of freedom and invariance is with respect to the observers built of other physical fields.  Here we can imagine inducing a set of ``physical coordinates'' based on local packets of long lasting separated objects that define a grid.  With the right time evolution parameterization, we would expect the form of the equations to be invariant with respect to such a coordinate set.  The overall class of equivalent solutions should allow for local field changes that induce independent observable current changes with the appropriate degrees of freedom for the observed dynamic freedom of the system.  In general, we only need observers to see the world with such symmetry (such as Lorentz) but it need not hold with respect to the coordinates.  As long as the constituent fields of the observers and the external reality ``covary'' together, then the observers see exactly the same thing.  Allowing such dynamics can enlarge the equivalence classes at the cost of a more complicated relationship between coordinates and observable reality.  

Generally we seek a quadratic free field lagrangian and then gauge and Lorentz invariant couplings between them.  The Dirac lagrangian is usually presented in the superficially symmetric form  
\begin{equation}\label{DiracLag}
\mathcal{L}_{D}=i\bar{\psi}\gamma^{\mu}\partial_{\mu}\psi-m\bar{\psi}\psi.
\end{equation}
The appearance of the $\gamma^{0}$ is displeasing if we are to interpret the $\mu$ indices as spacetime indices.  This particular form is often considered important because it gives a positive definite probability density.  In an ``emergent'' approach to quantum theory where the probabilities are defined by the evolution equations in a deterministic fashion, this is not important.  Probability will automatically be conserved by the normalization over the resulting paths that bifurcate the histories of recording devices and observers as indexed by the delocalized particle's coordinates \cite{Chafin} regardless of whether there is a ``nice'' operator that describes it.  More importantly, we need the eom of $\psi$ and $\psi^{*}$ to be consistent.  This dictates that the $\gamma^{0}$ appear in this expression.  By using a representation where $\gamma^{0}\gamma^{\mu}\gamma^{0}=\gamma^{\mu}$ the variations of the action give equivalent equations of motion.

To achieve a lagrangian that is manifestly invariant using this ``vector-plet'' interpretation we introduce an auxiliary field $\phi$ that, in flat space, can be chosen to be ${\psi}^{*}\gamma^{0}$.  For the usual Dirac equation this condition is propagated.  One should wonder if this will give a true isomorphism with physical results.  We are interested in the propagation of conserved quantities as mass, charge\ldots and some local phase information.  This brings us to a subtle point.  Even in nonrelativistic quantum mechanics, the ``reality'' of interacting particles is not completely given by the corresponding fields themselves.  This is most clearly observed in the AB effect.  Often this is viewed as an important example of topology and gauge in physics.  It is more simply understood as an expression of the electron current being not simply a function of the electron wavefunction alone.  A similar property is observed in the London skin depth in superconductors.  The only way an electron current can obtain rotational flow is through the vector field $\vec{A}$ or through the appearance of discrete vortices.  
The moral here is that angular momentum, among other conserved quantities, is defined by a collective set of fields so it makes no sense to associate with one particular particle.  ``Spin'' is now a kind of angular momentum that exists through the collective local reality of this new vector-plet graviton and two fermion spinplet fields.  By abandoning this usual concept of a spinor we will obtain an isomorphic theory that has significant generalizations.

\markright{Clifford Chafin. Gauge Freedom and Relativity}
\section{Bilinear modification}
\markright{Clifford Chafin. Gauge Freedom and Relativity}

To resolve the complications arising from the hidden $\gamma^{0}$ in the usual Dirac lagrangian, let us replace $\bar{\psi}$ with an associated yet independent field $\phi$ and see when it evolves in a consistent fashion when we simplify to the Dirac representation.  
Consider the Dirac-limiting lagrangian density we can choose using only the complex valued $\psi$, $\phi$ and $\gamma^{\alpha}$ (with $g^{\mu\nu}$ an implicit function of it) is of the form
\begin{equation}
\mathcal{L}=i\left(\phi_{a}\gamma^{\mu}_{ab}\partial_{\mu}\psi_{b}-\partial_{\mu}\phi_{a}\gamma^{\mu}_{ab}\psi_{b}\right)-2m\phi_{a}\psi_{a}.
\end{equation}
For constant $\gamma$'s chosen to be the Dirac representation, then variation $\delta\phi$ yields $i\gamma^{\mu}\partial_{\mu}\psi-m\psi=0$.  Variation by $\delta\psi$ yields $-i(\partial\phi)\gamma^{\mu}-m\phi=0$.  If we choose $\phi_{a}=\gamma^{0}_{ab}\psi^{*}_{b}$ then this is equivalent to the Dirac equation solution for $\phi$.  

When we consider the gauge equivalent states this introduces some additional considerations.  For example, if the support of $\psi$ and $\phi$ are disjoint then there is no net mass or current density.  Such a state is evidently a vacuum despite the nontrivial values of the functions and evolution equations.  Here we see that our notions of the physical meaning we attach to functions as describing the reality of a particle is less trivial than usual.  

So far we have not explicitly included any measure or metric and the action of $\nabla_{\mu}\gamma^{\nu}$ is ambiguous without it.  We can make formal definitions of these by using eqn.\ \ref{gamma} as a guide.  The pair of functions,
\begin{equation}
\left.\begin{array}{l}
\displaystyle
g^{\mu\nu} = ~-\rub \frac 14\,\mathrm{Tr}_{ac}^{\phantom{0}}\gamma^{(\mu}_{ab}\gamma^{\nu)}_{bc}\\[+7pt]
\displaystyle
g_{\mu\nu} = ~\mathrm{Inv}\left(-\rub \frac 14\,\mathrm{Tr}_{ac}^{\phantom{0}}\gamma^{(\mu}_{ab}\gamma^{\nu)}_{bc}\right)
\end{array}\quad \right\}
\end{equation}
 to define the metric in terms of $\gamma$ are evidently complicated when explicitly constructed but they do give us trial definitions for $g^{\mu\nu}(\gamma)$ and its inverse in terms of $\gamma^{\mu}$ that can specify a completely general metric field.  Another possible objections is that the form of $\gamma^{\mu}$ with indices raised as a contravariant object is opposite that of the covariant form that $A_{\mu}$ enters the lagrangian especially the interaction terms $q\bar{\psi}\gamma^{\mu}A_{\mu}\psi$ which gives us pause when considering the possibility of treating $\gamma^{\mu}$ and $A_{\mu}$ as analogous fields where no a priori metric exists.  

Since we are interested in a theory that includes electrons, positrons, photons and gravity with the electromagnetic and gravitational fields on an equivalent footing we will will need to make a further modification.  It will be convenient to let the natural form of $\gamma$ be a   lowered index object $\gamma_{\mu}$ and introduce a contravariant sister field $\lambda^{\nu}$ that generates $g^{\mu\nu}$ in the same fashion that $\gamma_{\mu}$ generates $g_{\mu\nu}$.  It is not automatic that these be inverse functions despite the suggestive notation but we will show that they do so in sufficiently low energy cases for a particular lagrangian.  We expect the following relations to be able hold in the flat space limit
\begin{equation}
\left.\begin{array}{ll}\label{id}
\displaystyle
g^{\mu\nu}\delta_{ac}=-\frac 12\,\{\lambda^{\mu},\lambda^{\nu}\}=-\lambda^{(\mu},\lambda^{\nu)}\\[7pt]
\displaystyle
g_{\mu\nu}\delta_{ac}=-\frac 12\,\{\gamma_{\mu},\gamma_{\nu}\}=-\gamma_{(\mu},\gamma_{\nu)}\label{g}
\end{array}\quad \right\}.
\end{equation}
It is very important to distinguish between this case, which arises in deriving the Klein-Gordon results that demonstrate causality for the Dirac components and the traced result.  The arbitrary metric field $g_{\mu\nu}(x)=-\frac 18${Tr}$\{\gamma_{\mu}(x),\gamma_{\nu}(x)\}$ can be defined in terms of $\gamma^{\mu}_{ab}(x)$'s but the untraced result for $g_{\mu\nu}(x)\delta_{ac}$ cannot.  This will be central to what follows.  

We like to have the metric appear explicitly in all the terms of the lagrangian for the reason it gives us something to vary in obtaining a conservation law for stress-energy.  One way to do this is is to use the lagrangian
\begin{equation}\label{electronL}
\mathcal{L}_{e}=i\left(g^{\mu\nu}\phi_{a}\gamma_{\mu:ab}\partial_{\nu}\psi_{b}-g^{\mu\nu}(\partial_{\mu}\phi_{a})\gamma_{\nu:ab}\psi_{b}\right)-2m\phi_{a}\psi_{a},
\end{equation}
where the colon separates spacetime from scalar indices.  We define $g^{\mu\nu}=-\frac 14 ${Tr}$\lambda^{(\mu},\lambda^{\nu)}$.  
The evolution equations are given by the variations $\delta\phi$
\begin{equation}
\left.\begin{array}{ll}
\displaystyle
i\left(g^{\mu\nu}\gamma_{\mu:ab}\partial_{\nu}\psi_{b}+g^{\mu\nu}\nabla_{\mu}(\gamma_{\nu:ab}\psi_{b})\right)-2m\psi_{a}=0\\[7pt]
\displaystyle
i\rub g^{\mu\nu}\gamma_{\mu:ab}\partial_{\nu}\psi_{b}+\frac 12\,i\rub g^{\mu\nu}(\nabla_{\mu}\gamma_{\nu:ab})\psi_{b}-m\psi_{a}=0
\end{array}\quad \right\}
\end{equation}
and $\delta\psi$
\begin{equation}
i\rub g^{\mu\nu}(\nabla_{\mu}\phi_{b})\gamma_{\nu:ba}+\frac 12\,i\rub g^{\mu\nu}\phi_{b}(\nabla_{\mu}\gamma_{\nu:ba})+m\phi_{a}=0
\end{equation}
so that $\phi$ evolves as $\psi$ with $m\rightarrow-m$ and $\gamma\rightarrow\gamma^{\mathrm{T}}$.\footnote{Note that this does not mean that the energy of the rest field is $m$ ($c=1$).  The energy is a function of the triple of fields $(\psi,\phi,\gamma)$ as we see next.}  

Since we are about to determine the motion of the conserved gauge invariant stress energy associated with the fields and it is deeply connected with geometry, we make a brief segue to derive this conserved quantity.  
A general action contains both a lagrangian and a measure that can be related to the metric
\begin{equation}
S=\int d^{4}x \,\mathcal{L}\sqrt{-g_{\cdot\cdot}} \ .
\end{equation}
Incorporating general relativity, the lagrangian density is generally written
\begin{equation}
\mathcal{L}=\frac{1}{2\varkappa}\,R(g)+\mathcal{L}_{\mathrm{fields}},
\end{equation}
where $\varkappa=8\pi G$ and the first term gives the Riemann curvature and the second gives the field terms that do not depend only on the metric.  The conservation laws arise from varying the metric $\delta g^{\mu\nu}$ from which we obtain
\begin{equation}
G^{\mu\nu}=8\pi G T^{\mu\nu}=-\varkappa \frac{-2}{\sqrt{-g^{\cdot\cdot}})^{-1}}\frac{\delta\mathcal{L}_{\mathrm{fields}}(\sqrt{-g^{\cdot\cdot}})^{-1}}{\delta g^{\mu\nu}}.
\end{equation}
Since $\nabla_{\mu}G^{\mu\nu}=0$ as an identity we have $\nabla_{\mu}T^{\mu\nu}=0$.  This is a local conservation law.  To obtain a global one we need a spacetime with persistent Killing vectors corresponding to continuous symmetries.  The action of gravity typically destroys these as global conservation laws, however, if $G\rightarrow0$ \textit{and} the initial data is chosen to be flat then these exist and persist so we have the usual global symmetric conservation laws.  This justifies this as a general method of deriving conservation laws with symmetric stress-energy tensors for fields on flat space when all the fields present are tensorial.  Of course, we expect any such conservation law to correspond to a symmetry.  In this case, we can vary the coordinates locally and this leaves the quantity $\mathcal{L}\sqrt{g_{\cdot\cdot}}$ invariant.  Since all the derivatives are covariant, we can replace a passive coordinate change on an open set with an active transformation of the metric field $g^{\mu\nu}$.  Varying $g^{\mu\nu}$ is therefore equivalent to a general small variation in the local coordinates.  Of course, we are considering these as fields on a flat background so that they change in a rather simple fashion relative to the coordinate changes and we should include a coordinate measure $\sqrt{-\eta}$ and this underlying space generates full set of ten conserved quantities (see \S3).

The (symmetric) stress tensor is usually defined by\footnote{Here we make the choice of taking the determinant with respect to the ``contravariant'' metric $g(\gamma^{\mu})$ in anticipation of later work.  This explains the power -1 this expression.}  
\begin{equation}
\begin{array}{rl}
\displaystyle
T_{\mu\nu}=&-\displaystyle\frac{2}{(\sqrt{-g^{\cdot\cdot}})^{-1}}\frac{\delta \left( \mathcal{L}_{\mathrm{fields}}\left(\sqrt{-g^{\cdot\cdot}}\right)^{-1}\right)}{\delta g^{\mu\nu}}\\[+12pt]
=& -2\displaystyle\frac{\delta \left( \mathcal{L}_{\mathrm{fields}}\right)}{\delta g^{\mu\nu}}+g_{\mu\nu}\mathcal{L}_{\mathrm{fields}}\\[+9pt]
=&2i\left(\phi_{a}\gamma_{(\mu:ab}\partial_{\nu)}\psi_{b}-(\partial_{(\mu}\phi_{a})\gamma_{\nu):ab}\psi_{b}\right)\\[+6pt]
&+\,g_{\mu\nu}\Bigl[\, i\,\bigl(g^{\alpha\beta}\phi_{a}\gamma_{\alpha:ab}\partial_{\beta}\psi_{b}\\[+6pt]
   &-\,g^{\alpha\beta}(\partial_{\alpha}\phi_{a})\gamma_{\beta:ab}\psi_{b}\bigr)-2m\phi_{a}\psi_{a}\Bigr] \\[+6pt]
= & 2i\left(\phi_{a}\gamma_{(\mu:ab}\partial_{\nu)}\psi_{b}-[\partial_{(\mu}\phi_{a}]\gamma_{\nu):ab}\psi_{b}\right),
\end{array}
\end{equation}
where we have varied with respect to $g^{\mu\nu}$ and assumed $\gamma_{\mu}$ is a field independent of it in anticipation of $g^{\mu\nu}$ being a function of $\lambda^{\mu}$.  

We can similarly examine the continuous symmetry given by the globally constant phase changes $\psi\rightarrow e^{i\theta}\psi$ and $\phi\rightarrow e^{-i\theta}\phi$ to get the conserved current
\begin{equation}
j^{\nu}=2i\rub g^{\mu\nu}\phi_{a}\gamma_{\mu:ab}\psi_{b}
\end{equation}
so that $\nabla_{\nu}j^{\nu}=0$.  Here we see this current also depends on all three fields so that the vanishing of any one of them on a region necessitates the entirety of the physical reality vanish.

We will now consider the implications of packet motion given these two conservation laws.  Firstly, when we say\linebreak 
``packet'' we are not referring to a packet of localized $\psi$ or $\phi$ as much as a localized region where the reality associated with these fields through $T_{\mu\nu}$ and $j^{\mu}$ are nonzero.  Let us also consider a packet that is devoid of internal stress and rotation and where the pressure is minimal.  For such a packet with sufficiently uniform interior we can average over the current to give 
$\braket{j^{\mu}} \approx m v^{\, \mu}$ 
where $m^{2}$ is the averaged $g_{\mu\nu}j^{\mu}j^{\nu}$ density and, assuming the packet preserves its structure as it moves, $v^{i}$ is the local coordinate velocity of the packet.  We can then define $v^{0}$ by the relation $g_{\mu\nu}v^{\,\mu}v^{\nu}=-1$.  The conservation law tells us that $\rho$ is conserved.  $v^{\,\mu}$ is well defined to the extent packet motion is so.  

From $\braket{T^{\mu 0}}$ we can define a velocity $u$ that carries the energy in a localized packet so that $\braket{T^{\mu 0}}\approx m' u^{(\mu}u^{0)}$.  Since a vanishing of the current on a region implies vanishing of stress-energy as well we have that $v=u$ and that $\braket{T^{\mu 0}} \approx m^{'(\mu}v^{0)}=\alpha m^{(\mu} v^{0)}$.  Since there are no internal stresses, 
$\braket{T^{\mu\nu}} \approx\alpha m v^{\,\mu} v^{\nu}$.    
By combining  these expressions we derive that these ``macroscopic'' variables are 
\begin{equation}
\left.\begin{array}{ll}
v^{\nu}\!\!\!&=\displaystyle\frac{\braket{T^{\mu\nu}}}{\alpha \braket{j^{\mu}}}\\[+10pt]
m\!\!\!&=\alpha^{2}\displaystyle\frac{\braket{j^{\mu}} \braket{j^{\nu}}}{\braket{T^{\mu\nu}}}
\end{array}\quad \right\},
\end{equation}
where these are actually several equations (repeated indices are not summed) that are all equal by the conditions above.  

Now consider the parcel averaged stress-energy conservation law.  Applying $\nabla_{\mu}j^{\mu}=0$ we have
\begin{equation}
\begin{array}{ll}
\braket{\nabla_{\mu}T^{\mu \nu}} & = \braket{\nabla_{\mu}(j^{\mu}v^{\nu})} \\[5pt]
&= \braket{(\nabla_{\mu}j^{\mu})v^{\nu}+j^{\mu}\nabla_{\mu}v^{\nu}} \\[5pt]
&=m' \braket{\nabla_{v}v} = 0,
\end{array}
\end{equation}
which indicates the gauge invariant aspects (i.e.\ the reality) of the parcel follows geodesic motion.  This is not entirely surprising given that it is known that the conservation laws generally dictate that classical particles follow geodesics though the proofs are generally quite difficult \cite{Ehlers}.  The ``geodesics'' here are generally curved paths in our underlying coordinate space but appear as geodesics in the geometry most apparent to observers.  

In the next section for a theory of ``lepto-electro-gravity'' we have two covariant gauge fields and one contravariant one.  These have trivial transformation laws in the flat background coordinates but we maintain this distinction because it seems more relevant for observers.  In this sense we think of it as a ``2+1'' theory.  One contravariant field is always necessary to match the covariant derivatives that must arise in any differential equation.  The electron field is described by a $(\phi,\psi)$ pair of fields that embody its reality with a very large gauge group and the meaning of the reality they describe depends not only on the metric but the covariant gravity field $\gamma_{\mu}$.  We will see that these have properties that are distinct from the positive energy positrons so we will require another pair of fields for their description.  Along the way we will introduce a lagrangian that exists as a purely polynomial expression and removes the need for complicated nonanalytic measures and rational inverse matrix functions.

\markright{Clifford Chafin. Gauge Freedom and Relativity}
\section{Electro-gravity lagrangian}
\markright{Clifford Chafin. Gauge Freedom and Relativity}

Here we seek a lagrangian that encompasses electrons,\linebreak
 positrons, electromagnetism and gravity and seek to have\linebreak 
equations that are polynomial rather than complicated rationals that arise from the operation of taking the inverse of the metric.  For this reason we define the function $g:\mathcal{V}\rightarrow \mathcal{T}$ where $\mathcal{V}$ is the set of vector-plet objects $\lambda^{\mu}_{ab}$ and $\gamma_{\mu:ab}$ and $\mathcal{T}$ is the set of corresponding contravariant or covariant 2-tensors $g^{\mu\nu}$ and $g_{\mu\nu}$ respectively.  Specifically,
$$g(A,B)=-\frac 18\,\mathrm{Tr}(AB+BA).
$$  
We will establish a lagrangian that gives Dirac particle motion in the flat space limit, electromagnetism and a form for GR that gives a simple parallel between the motion of the gravitational fields, $\gamma_{\nu}$ and the electromagnetic ones $A_{\nu}$ that allows gravity to obtain the nonlinear ``geometric'' features of~GR.  

Since we are interested predominantly in positive energy solutions we will need to introduce a separate action term $\Lambda_{p}$ for positrons that have positive mass but a reversal of sign of the charge in the coupling.  We can write the lagrangian for the covariant gravitational field $\gamma$ by substitution into the Einstein-Hilbert lagrangian.  Alternately, we can choose it to have a similar form of the action $\Lambda_{g}'$ as the other vector potential $\Lambda_{A}$ and the coupling terms $\Lambda_{e\lambda A}$, $\Lambda_{p\lambda A}$ will involve both the contravariant gravitational field $\lambda$ and the vector potential.  Finally, there will need to be some way for the covariant and contravariant gravitational fields to relate to one another.  This will be accomplished by a Higgs-like interaction term $\Lambda_{c}$.  
The general action is then defined as
\begin{equation}
\begin{array}{rl}\label{S}
S=&\int d^{4}x \mathcal{L}\sqrt{-g}= \int d^{4}x ~\Lambda\\[+3pt]
=&\int d^{4}x~(\Lambda_{g}+\Lambda_{\lambda}+\Lambda_{A}+\Lambda_{e}+\Lambda_{p}\\
&~~~~~~~~~~~+\Lambda_{e\lambda A}+\Lambda_{p\lambda A}+\Lambda_{c}),
\end{array}
\end{equation}
where we will define $\Lambda_{\lambda}$ shortly.  

Since the measure is a nonanalytic function of the metric but this is not retained in the usual equations of motion.  We will find that this is also true here.  For reasons as above we use the $\lambda$ fields in defining the measure.  

The electron part of the action is given by the substitutions
\begin{equation}
\begin{array}{rl}\label{Le}
\Lambda_{e}=& \mathcal{L}_{e}\left(\!\sqrt{g^{\cdot\cdot}(\lambda)}\right)^{-1}\\[+3pt]
=&\Bigl[\,i\left(g^{\mu\nu}(\lambda)\phi_{a}\gamma_{\mu:ab}\nabla_{\nu}\psi_{b}-g^{\mu\nu}(\lambda)(\nabla_{\mu}\phi_{a})\gamma_{\nu:ab}\psi_{b}\right)\\
  &-\,2m\phi_{a}\psi_{a}\,\Bigr]\,\bigl(\!\sqrt{g^{\cdot\cdot}(\lambda)}\bigr)^{-1}
\end{array}\!\!\!\!
\end{equation}
where we have, harmlessly, replaced the ordinary with covariant derivatives since the act on spinplet objects which are essentially scalars.  Variation with the measure present allows their action on higher tensors to give the appropriate covariant connection terms.  This is one indication of how the physics itself can generate the geometric aspects of gravity rather than imposing it by fiat in the formulation of the theory's foundations.

The positron portions of the lagrangian is of the same form as $\Lambda_{e}$ but with a different pair of fields $\tilde{\phi}, \tilde{\psi}$.  The distinction comes in the form of the interaction terms.  The usual minimal coupling prescription gives 
\begin{equation}
\left.\begin{array}{ll}
\Lambda_{e\lambda A}&\!\!\!\!=-\,q\rub\phi_{a}\lambda^{\mu}_{ab}A_{\mu}\psi_{b}\\[+7pt]
\Lambda_{p\lambda A}&\!\!\!\!=+\,q\rub\tilde{\phi}_{a}\lambda^{\mu}_{ab}A_{\mu}\tilde{\psi}_{b}
\end{array}\quad \right\}.
\end{equation}
It is only the sign of the charge in the interaction terms that distinguishes positrons from electrons and it only appears in the couplings.

The gravitational part of the action can be defined by a simple extension of the Einstein-Hilbert action
\begin{equation}\label{EH}
\Lambda_{g}=\frac{1}{2\varkappa}\,R\left(g_{\mu\nu}(\gamma),g^{\mu\nu}(\lambda)\right)\left(\!\sqrt{g^{\cdot\cdot}(\lambda)}\right)^{-1}. 
\end{equation}
$R$ is defined in terms of $g_{\mu\nu}(\gamma)$, $g^{\mu\nu}(\lambda)$ and the connections implicit in the expression are defined by 
\begin{equation}
\Gamma^{\alpha}_{\mu\nu}=\frac 12\,g^{\alpha \sigma}(\lambda)\left(g_{\mu\sigma,\nu}(\gamma)+g_{\sigma\nu,\mu}(\gamma)-g_{\mu\nu,\sigma}(\gamma)\right)
\end{equation}
and their derivatives.  
We expect that some induced con-\linebreak 
straints force $g(\gamma)g(\lambda)=\delta$.  To have this done as a result of field interactions we  exploit a ``Higgs-ish'' mechanism with the coupling term
\begin{equation}
\Lambda_{c}=M\left|\,g_{\mu\nu}(\gamma)\,g^{\nu\rho}(\lambda)-\delta_{\mu}^{\rho}\right|^{2}
\end{equation}
for a sufficiently large mass $M$.  When the energies in the other terms are much smaller this drives the relation between $\gamma$ and $\lambda$ to hold so that the solutions become ``geometric.''  Specifically, while it is easy to enforce causality if all evolution fields obey some equation such as $g^{\mu\nu}\partial_{\mu}\partial_{\nu}\phi+\ldots$ where $g^{\mu\nu}$ is a metric with signature $+2$, the geometric case indicates that slowly spreading packets in regions of slowly varying spacetime move along geodesics.  When such a relation holds our lagrangian has a form that can be interpreted as coordinate invariant in that the derivatives act on the tensor fields with covariant derivatives with the $\Gamma$s induced by the metric $g_{\mu\nu}=-4^{-1}\mathrm{Tr}\gamma_{(\mu}\gamma_{\nu)}$.  In the next section we will see that we can also interpret the system to live on a flat background and derive global conservation laws.  

The other gauge fields all come from lagrangians that\linebreak 
have electromagnetic form $F^{\mu\nu}F_{\mu\nu}$ where $F_{\mu\nu}=\partial_{\mu}A_{\nu}-\partial_{\nu}A_{\mu}$.  Specifically,
\begin{equation}
\begin{array}{rl}
\Lambda_{A}= & g^{\mu\alpha}(\lambda)g^{\nu\beta}(\lambda)(\partial_{\mu}A_{\nu}-\partial_{\nu}A_{\mu})\\[5pt]
             &  \times\, (\partial_{\alpha}A_{\beta}-\partial_{\beta}A_{\alpha})\left(\!\sqrt{-g^{\cdot\cdot}(\lambda)}\right)^{-1}.
\end{array}
\end{equation}
It is not necessary to use covariant derivatives here since antisymmetry cancels them.  
For example, we model the action contribution from the ``dual field'' $\lambda$  as
\begin{equation}
\begin{array}{rl}
\Lambda_{\lambda}= &\epsilon~g^{\mu\alpha}(\lambda)\rub g^{\nu\beta}(\lambda)\,\mathrm{Tr}(\partial_{\mu}\tilde\lambda_{\nu}-\partial_{\nu}\tilde\lambda_{\mu})\\[5pt]
                   &  \times\,(\partial_{\alpha}\tilde\lambda_{\beta}-\partial_{\beta}\tilde\lambda_{\alpha}) \left(\sqrt{-g(\lambda)}\right)^{-1},
\end{array}
\end{equation}
where $\tilde\lambda_{\mu}=g_{\mu\nu}(\gamma)\lambda^{\nu}$.\footnote{We distinguish this field with a tilde because of the earlier convention that these are all tensor indices under the underlying flat space metric so that ``lowering'' an index with $g$ must be new field to not be ambiguous.}
where we have chosen the constant $\epsilon$ to be small so that the dynamics can be dominated by $\gamma$ and the constraints induced by the Higgs-like term.  

For a function $F_\mu(g_{\mu\nu}(\gamma))$ the variation under $\delta\gamma_{\nu}$ gives 
\begin{equation}
\delta F_{\mu}=\frac{\delta F}{\delta g_{\mu\nu}}\,\delta\gamma_{\nu}
\end{equation}
and similarly for $\delta\lambda$.  
Variation of $\Lambda_{g}$ by $\delta\lambda$ gives
\begin{equation}
\frac{1}{2\varkappa}\left( R_{\mu\nu}-\frac 12\,R g_{\mu\nu} \right)\delta\lambda^{\nu}
\end{equation}
or
\begin{equation}
G_{\mu\nu}\delta\lambda^{\nu}=\varkappa T_{\mu\nu}\delta\lambda^{\nu},
\end{equation}
where $T_{\mu\nu}$ is the stress-energy tensor for all the actions terms other that $\Lambda_{g}$.  We have implicitly assumed that we are in a low enough energy regime and the initial data includes no ``waves'' of $\lambda$ so that the contributions of $\Lambda_{\lambda}$ can be ignored.  Since the $\gamma$'s contain gauge freedom that is independent of coordinate changes so that we can choose any $\gamma_{\mu}$ that give the same $g_{\mu\nu}(\gamma)$ field, this requires
\begin{equation}
G_{\mu\nu}=\varkappa T_{\mu\nu}\,.
\end{equation}

\markright{Clifford Chafin. Gauge Freedom and Relativity}
\section{Conservation laws}
\markright{Clifford Chafin. Gauge Freedom and Relativity}

We can argue the whole structure exists on a flat background though this is just a convenient artifice among many.  It is however a very convenient one.  The appearance of geometric evolution via the additional $\Gamma$ factors that make the derivatives seem ``covariant'' with respect to some induced geometry of these fields is an emergent byproduct of the kind of couplings present.  It should be noted that these $\Gamma^{\alpha}_{\beta\gamma}$ factors are actual $\eta$-tensors on the background space instead of affine connections.  Of course, we still need to know if our equations can be evolved for arbitrary times using this point of view.  Some discussion of this, especially in the case of black hole formation is given in \cite{Chafin-BH}.  For now we assume that this is unlimited however, although other methods have attempted to justify working on a flat background \cite{Lasenby} it is a delicate process to have this make sense as gravitational collapse ensues due to the trend of the equations to become ill conditioned here.  One should not be overly comfortable with formalism in this case.  A method to handle evolution on the large regions of nearly degenerate metric using conservation laws is proposed in \cite{Chafin-BH}

The flat background has a natural set of Killing vectors that give global conservation laws.  To elucidate this consider the lagrangian written in terms of ordinary derivatives and make the modification 
by defining $g(\gamma)=h(\gamma)\circ\eta$
\begin{equation}\label{doublemeasure}
\Lambda=\mathcal{L}\sqrt{-g}\rightarrow \Lambda\sqrt{h}\sqrt{-\eta} \ .
\end{equation}

All actions on tensors induced by $\eta$-background coordinate transformations are of the form
\begin{equation}
\partial_{\mu}A^{\alpha}\rightarrow\nabla_{\mu}(\eta)A^{\alpha}=\partial_{\mu}A^{\alpha}-\Gamma^{\alpha}_{\mu\nu}(\eta)A^{\nu}
\end{equation}
and so forth,
where $\eta$ is a metric (in any coordiates) that can be varied about the flat space case.  Any covariant derivatives $\nabla_{\mu}(g)$ in terms of the metric induced connections are reinterpreted as formal couplings through $\Gamma(g)$ and the $\partial_{\mu}$ are converted by this prescription.  We see a problem with eqn.\ \ref{doublemeasure} is that it is not invariant under general $\eta$-space coordinate transformations due to the factor $\sqrt{-g}$.  It is, however, invariant under the isometries of flat spacetime that we use to generate global conservation laws.

Since the flat space contains a full set of ten Killing vectors we have a set of conserved global quantities that now includes the gravitational fields of the form
\begin{equation}
\partial_{\mu}T^{'\mu\nu}=0
\end{equation}
with the Killing (co)vector fields $p_{\nu}=\hat{\omega}_{\nu}$, $M_{ijk}=\epsilon_{ijk}x_{j}\hat{\omega}_{k}$ and $b_{i}=x_{0}\hat{\omega}_{i}+x_{i}\hat{\omega}_{0}$.  The globally conserved quantities in these coordinates are
\begin{equation}
\left.\begin{array}{ll}
P^{\nu}=&\!\!\!\!\int d^{3}x ~p_{\mu}T^{'\mu\nu}\\[+4pt]
J^{i}=&\!\!\!\!\int d^{3}x ~M^{i}_{jk}T^{'ij}\\[+4pt]
C^{j}=&\!\!\!\!\int d^{3}x ~b_{i}T^{'ij}
\end{array}\quad \right\}.
\end{equation}

\markright{Clifford Chafin. Gauge Freedom and Relativity}
\section{Conclusions}
\markright{Clifford Chafin. Gauge Freedom and Relativity}

The notions of invariance from differential geometry and invariance theory are imported into physics in a fashion that ranges from formal to ad hoc.  Surprisingly, they have not been reconsidered from the more physical point of view that all configurations that are indistinguishable to observers built of the fields themselves should form the most general equivalence class of systems.  This enlarged meaning of ``gauge'' requires some underlying structure.  We have shown that many of the usual objections to a flat background can be overcome and that this allows the fields to have very simple transformation laws and a large set of conservation laws with respect to this flat background.  The observers can then perceive a curved space with all its mathematical complexity as emerging from the nature of nonlinear and multilinear coupling among fields.  Importantly, there is a classical lagrangian with a Higgs-like term that causes there to be such a strongly nonlinear and geometric theory of gravity to arise from the perspective of such observers at low energy.  

An interesting by-product of this approach is that the apparent co and contravariant properties of the fields in the\linebreak 
``physical coordinates'' induced by objects for the observers obtain their transformation properties by the equations of motion not by a by-fiat assignment.  This is another aspect of ``geometry'' that is determined by the physics itself.  At high enough energies we expect this geometric association to fail and nonmetric features to become evident to the observers.  In this case the induced constraints fail and evolution becomes potentially more difficult.  One suggestion is that such a situation allows inconsistent light cone structures to be induced for different fields and that some intersection of these gives the proper causal structure for these fields when they are interacting.  

The bilinear extension of the Dirac equation and promotion of the $\gamma$ matrices to dynamical fields introduced a number of concerns related to positive definiteness of energy and probability and causality of the equations of motion.  The latter has been verified for packets using gauge invariant functions of the fields.  
The former is seen to be not essential since these quantities, while rigidly conserved, are not necessarily the physical ones an observer perceives since they are derived from background coordinate symmetries.  The probability function may be a nontrivial function of the fields in the case of gravity but normalization is assured in any theory of emergent measurement such as decoherence.

There are undoubtably many inequivalent such theories with the same low energy limit so we have presented only one of probably many such solutions.  From here it is unclear how to extend this classical theory to a quantum one.  The couplings are such that they determine the local notion of causality and it is not clear when or how well a perturbative scheme, which is generally built on free fields solutions, will work in the many body case.  This is a direction for future~work.


%
\begin{flushright}\footnotesize
Submitted on November 4, 2014 / Accepted on November 24, 2014
\end{flushright}

\vspace*{-6pt}
\centerline{\rule{72pt}{0.4pt}}
}


\newpage

\begin{thebibliography}{99}\footnotesize



\bibitem{Ma} Marquet P.  Lichnierowicz's theory of spinors in General Relativity:
the Zelmanov approach. \textit{The Abraham Zelmanov Journal}, 2012 v.\,5, 117--133. 

\bibitem{Bjorken}
Bjorken J.D., Drell S.D.  Relativistic Quantum Mechanics and Relativistic Quantum Fields. McGraw-Hill, 1965.

\bibitem{Peskin}  Peskin M.E. and Schroeder D.V.   An Introduction to Quantum Field Theory, Ed.\ Westview, Boulder, 1995.

\bibitem {Weinberg} Weinberg S. The Quantum Theory of Fields.
Volumes I and II. Cambridge University Press, Cambridge, 1995.

\bibitem{Hestenes}
Hestenes D. 
A unified language for Mathematics and Physics \& Clifford algebra and the interpretation of quantum mechanics. In: Clifford Algebras and Their Applications in Mathematics and Physics. J.S.R. Chisholm \& A.K. Common, eds., Reidel, Dordrecht, 1986, pp. 1--23 and pp. 321--346.


\bibitem{Chafin-nonloc} Chafin C. Automorphism Induced Nonlocal Conservation Laws. math-ph/arXiv: 1407.6782.



\bibitem{Chafin}
Chafin C. 
The Quantum State of Classical Matter I: Solids and Measurements. quant-ph/arXiv:1308.2305.


\bibitem{Mach}
Mach E.  The Science of Mechanics; a Critical and Historical Account of its Development. LaSalle, IL: Open Court Pub. Co., 1960.

\bibitem{Einstein}  Einstein A. On the electrodynamics of moving bodies. 
\textit{Annalen der Physik}, 1905, v.\,17, 891--921.



\bibitem{MTW} Misner C.W., Thorne K.S. and  Wheeler J.A.   Gravitation. Freeman, 1973. 

\bibitem{Oppenheimer} Oppenheimer J.R. and Snyder H.  On Continued Gravitational Contraction. \textit{Physical Review}, 1939, v.\,56(5), 455--459.

\bibitem{Chafin-BH}
Chafin C. Globally Causal Solutions for Gravitational Collapse. gr-qc/arXiv:1402.1524.


\bibitem{Schweber}
Schweber S.S.
An Introduction to Relativistic Quantum Field Theory. Harper and Row,
1962.

\bibitem{Jackson}
Jackson J.~D. Classical Electrodynamics.  Wiley, New York, 1962.


\bibitem{Rohrlich}
Rohrlich F. Classical Charged Particles.
World Scientific,
2007.





\bibitem{LL} Landau L. and Lifshitz E. M. The Classical Theory of Fields. Pergamon, Oxford, 1979. 

\bibitem{Lasenby}
Lasenby A. Doran C. and Gull S. 
Gravity, gauge theories and geometric algebra. 
  \textit{Philosophical Transactions of the Royal Society}, London, 1998, v.\,356, 487.
  



\bibitem{Ehlers}
Ehlers J. and Geroch R. Equation of motion of small bodies in relativity. \textit{Annals of Physics}, 2004, v.\,309, 232--239.








\end{thebibliography}
\end{document}